\documentclass[twocolumn,superscriptaddress,floatfix,preprintnumbers,amssymb,amsmath]{revtex4}
\usepackage{graphicx}
\usepackage{dcolumn}
\usepackage{bm}
\usepackage[latin1]{inputenc}
\usepackage[mathscr]{eucal}
\usepackage{epsfig}
\usepackage{float}
\usepackage{pdfpages}

 \usepackage{pgfplots}
 \usepackage{tikz}

\newcommand{\varpm}{\mathbin{\vcenter{\hbox{\oalign{\hfil$\scriptstyle+$\hfil\cr\noalign{\kern-.3ex} $\scriptscriptstyle({-})$\cr}}}}}
\newcommand{\varmp}{\mathbin{\vcenter{\hbox{\oalign{\hfil$\scriptstyle-$\hfil\cr\noalign{\kern-.3ex} $\scriptscriptstyle({+})$\cr}}}}}

\newcommand{\itof}{i \to f}
\newcommand{\ftoi}{f \to i}

\newcommand{\Vitof}[1]{{#1}_{\itof}}
\newcommand{\Vftoi}[1]{{#1}_{\ftoi}}
\newcommand{\VxoiLR}[1]{{^{L(R)}{#1}^{y}_{x:0 \to 1}}}
\newcommand{\VxioLR}[1]{{^{L (R)}{#1}^{y}_{x:1 \to 0}}}
\newcommand{\VyoiD}[1]{{^D{#1}_{x}^{y:0 \to 1}}}
\newcommand{\VyioD}[1]{{^D{#1}_{x}^{y:1 \to 0}}}

\newcommand{\VxoiyoL}[1]{{^{L}{#1}^{y = 0}_{x: 0 \to 1}}}
\newcommand{\VxoiyiL}[1]{{^{L}{#1}^{y = 1}_{x: 0 \to 1}}}

\newcommand{\VxoiyoR}[1]{{^{R}{#1}^{y = 0}_{x: 0 \to 1}}}
\newcommand{\VxoiyiR}[1]{{^{R}{#1}^{y = 1}_{x: 0 \to 1}}}

\newcommand{\VxoyoiD}[1]{{^{D}{#1}^{y: 0 \to 1}_{x = 0}}}

\newcommand{\VxoyioD}[1]{{^{D}{#1}^{y: 1 \to 0}_{x = 0}}}

\newcommand{\VxpL}[1]{{^{L}{#1}^{y}_{x \to x + 1}}}
\newcommand{\VxpR}[1]{{^{R}{#1}^{y}_{x \to x + 1}}}
\newcommand{\VxpLR}[1]{{^{L(R)}{#1}^{y}_{x \to x + 1}}}
\newcommand{\VypD}[1]{{^{D}{#1}^{y \to y + 1}_{x}}}
\newcommand{\VxmL}[1]{{^{L}{#1}^{y}_{x + 1 \to x}}}
\newcommand{\VxmR}[1]{{^{R}{#1}^{y}_{x + 1 \to x}}}
\newcommand{\VymD}[1]{{^{D}{#1}^{y + 1 \to y}_{x}}}

\newcommand{\Energy}{E}
\newcommand{\Eitof}{\Vitof{\Energy}}
\newcommand{\ExoiLR}{\VxoiLR{\Energy}}
\newcommand{\ExioLR}{\VxioLR{\Energy}}
\newcommand{\EyoiD}{\VyoiD{\Energy}}
\newcommand{\EyioD}{\VyioD{\Energy}}

\newcommand{\Rate}{\Gamma}

\newcommand{\Ritof}{\Vitof{\Rate}}
\newcommand{\Rftoi}{\Vftoi{\Rate}}

\newcommand{\RxoiyoL}{\VxoiyoL{\Rate}}
\newcommand{\RxoiyiL}{\VxoiyiL{\Rate}}

\newcommand{\RxoiyoR}{\VxoiyoR{\Rate}}
\newcommand{\RxoiyiR}{\VxoiyiR{\Rate}}

\newcommand{\RxoyoiD}{\VxoyoiD{\Rate}}

\newcommand{\RxoyioD}{\VxoyioD{\Rate}}

\newcommand{\Current}{J}
\newcommand{\Jitof}{\Vitof{\Current}}

\newcommand{\JxpL}{\VxpL{\Current}}
\newcommand{\JxpR}{\VxpR{\Current}}
\newcommand{\JypD}{\VypD{\Current}}

\begin{document}

\title{Thermodynamics and efficiency of an autonomous on-chip Maxwell's demon}

\author{Aki Kutvonen}
\affiliation{COMP Center of Excellence, Department of Applied Physics,
Aalto University School of Science, P.O. Box 11000, FI-00076 Aalto, Espoo, Finland}
\author{Jonne Koski}
\affiliation{Low Temperature Laboratory, Department of Applied Physics, Aalto University School of Science, P.O. Box 13500, FI-00076 Aalto, Espoo, Finland}
\author{Tapio Ala-Nissila}
\affiliation{COMP Center of Excellence, Department of Applied Physics,
Aalto University School of Science, P.O. Box 11000, FI-00076 Aalto, Espoo, Finland}
\affiliation{Department of Physics, Box 1843, Brown University, Providence RI 02912-1843, U.S.A.}

\date{September 14, 2015}

\maketitle

\textbf{
In his famous letter in 1870, Maxwell describes how Joule's law can be violated "only by the intelligent action of a mere guiding agent", later coined as Maxwell's demon by Lord Kelvin. In this letter we study thermodynamics of information using an experimentally feasible Maxwell's demon setup based a single electron transistor capacitively coupled to a single electron box, where both the system and the Demon can be clearly identified. Such an engineered on-chip Demon measures and performes feedback on the system, which can be observed as cooling whose efficiency can be adjusted. We present a detailed analysis of the system and the Demon, including the second law of thermodynamics for bare and coarse grained entropy production and the flow of information as well as efficiency of information production and utilization. Our results demonstrate how information thermodynamics can be used to improve functionality of modern nanoscale devices.
}

Recent development of stochastic thermodynamics has extended the traditional macroscopic theory to small scales and non-equilibrium processes beyond linear response \cite{Jarzynski2011,Seifert2012,Bustamante2005,Collin2005}.  Information thermodynamics \cite{Parrondo2015,Sagawa2010,Sagawa2012,Horowitz2014,Barato2014}, which additionally considers processes that include information, measurement, and feedback, allows quantified studies on problems such as Maxwell's demon \cite{Leff2003}. The Demon is known as an object that acquires microscopic information of a system and applies feedback to decrease its entropy while, to retain the second law of thermodynamics, generates at least an equal amount of entropy. The emergence of nanotechnology has given rise to various theoretical proposals \cite{Horowitz2013, Mandal2013, Strasberg2013, Averin2011, Barato2013} as well as experimental realizations \cite{Toyabe2010, Koski2014,Parrondo2015,Koski2014b,Roldan2014,Koski2015} of a Maxwell's demon. The most recent studies in the field consider autonomous Demons - setups containing both the system measured and the Demon such that both the measurement and feedback are performed internally and no microscopic information needs to exit the system \cite{Horowitz2014,Strasberg2013,Barato2014,Mandal2013,Shiraishi2014,Ito2015}.

Recently it has been experimentally shown that an autonomous Maxwell's demon \cite{Koski2015} device based on single electron tunneling at low temperatures \cite{Averin1986, Lafarge1991,Buttiker1987,Pekola2013,Averin2011} can produce negative entropy in form of cooling its environment. More precicely, in the setup, a single electron transistor (SET) \cite{Kastner1992}, acts as the system to be measured, while the measurement and feedback is performed internally based on Coulomb interaction by a capacitively coupled single electron box, which acts as the Demon. The device has a limited number of relevant degrees of freedom, clear separation of different time scales, and well defined and measurable energy scales making it particularly suitable for studying dissipation at microscopic scales. In addition the device only requires fixed external voltage sources and a sufficiently low bath temperature to produce apparent negative entropy. The tunneling rates are not controlled externally during the operation. Here we study the role of information in the operation of the device in detail and show that by adjusting the properties of the Demon, the system's performance as a nanoscale cooling machine, including its efficiency, can be analyzed and tuned with thermodynamics of information.

\begin{figure*}{}
    \includegraphics[width=0.55\textwidth]{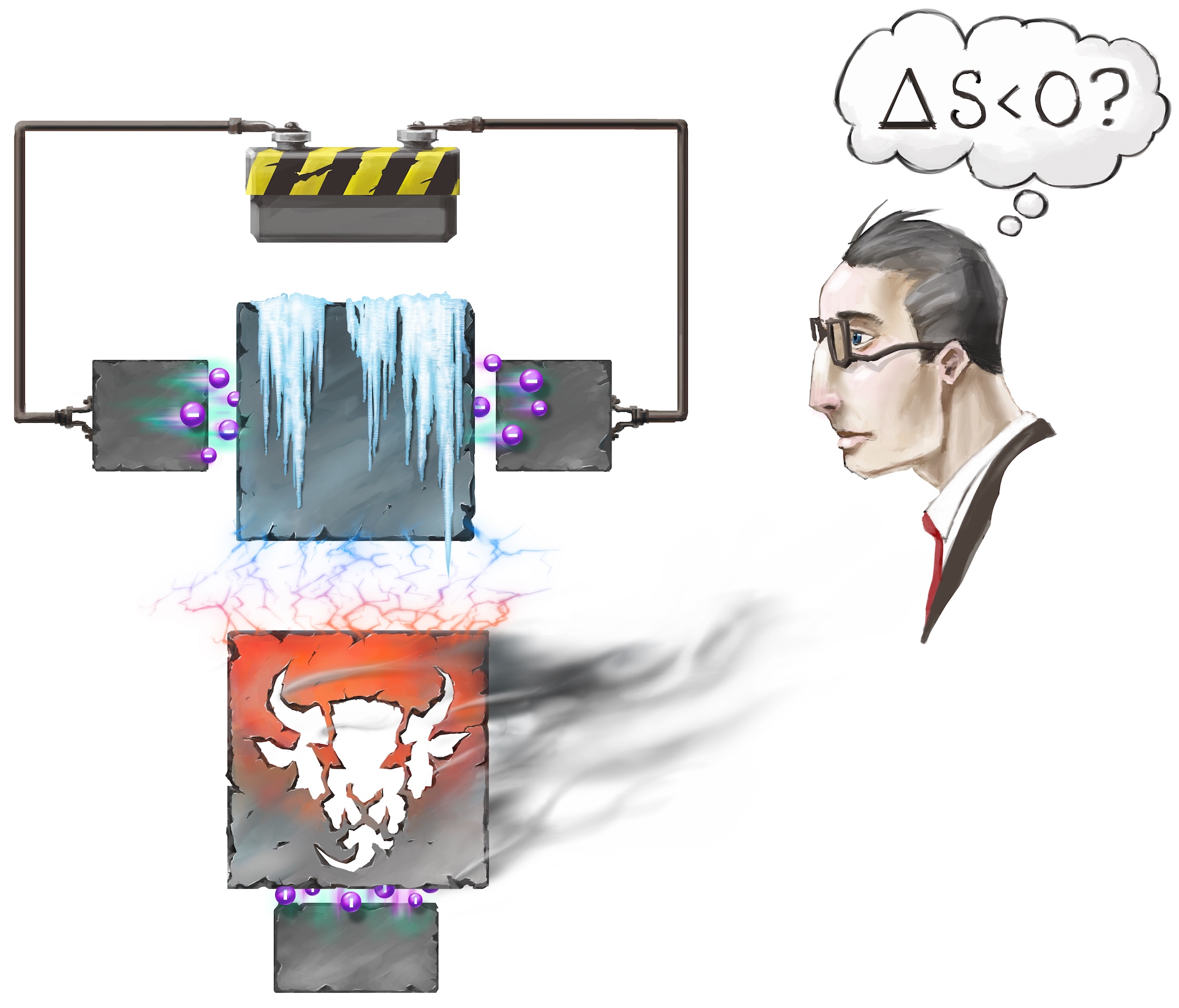}
       \raisebox{0.7cm}{\includegraphics[width=0.4\textwidth]{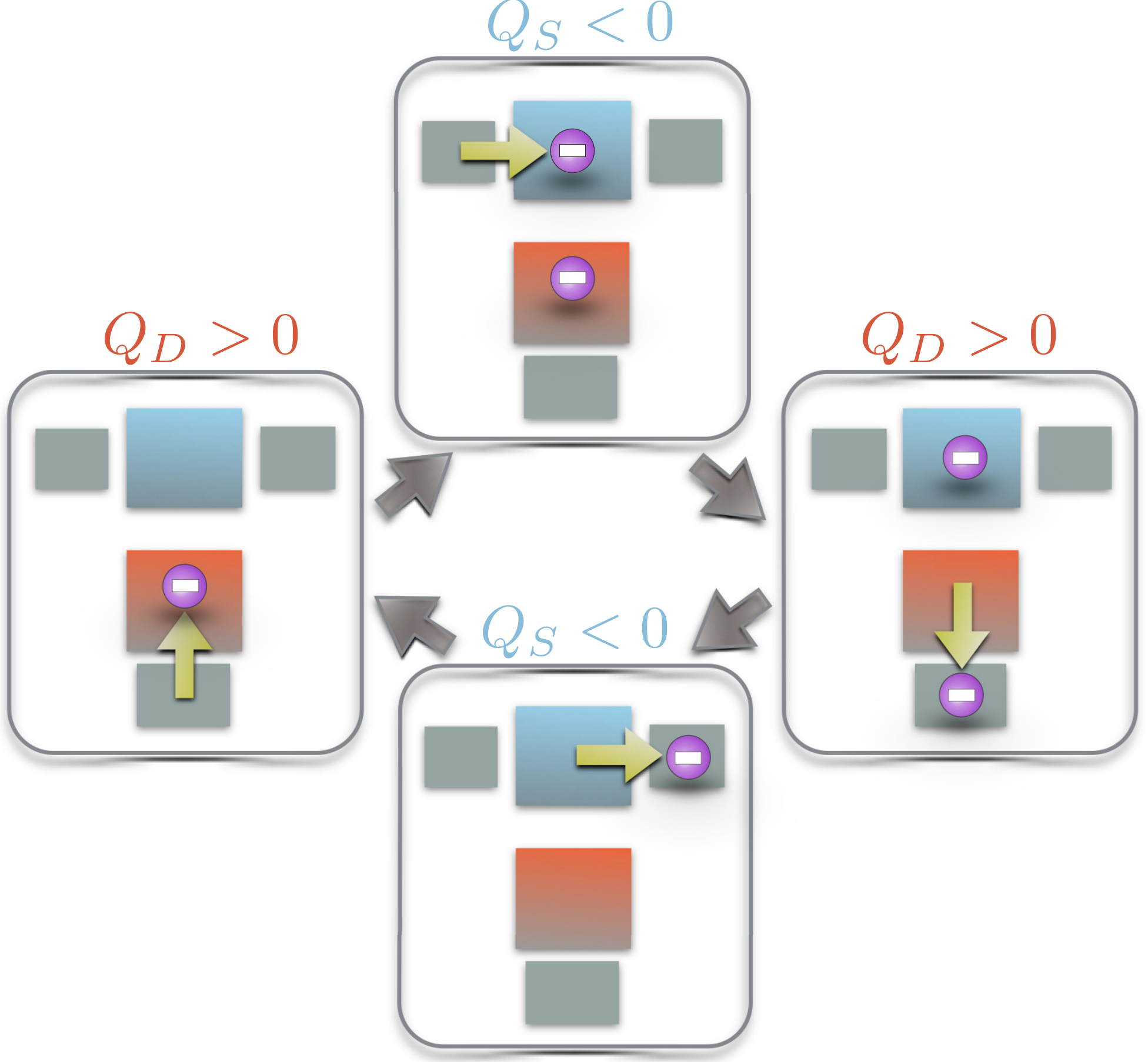}}
    \caption{\textbf{Schematic of the setup and the cooling cycle.} (a): A schematic picture of voltage biased SET capacitively coupled to an SEB detector, which acts as the Demon in the setup. Without seeing the Demon, the observer sees the SET system cooling even though the current runs through it. This would be a violation of Joule's law and second law of thermodynamics. However, the second law is retained by the heat dissipation in the Demon. Image by Heikka Valja. (b): The cooling cycle and dissipation in each step of the cycle. System tunneling events use thermal fluctuations to move the electron against the energy barrier. These events, illustrated in up and bottom images are accompanied by negative dissipation and cooling of the system. The Demon tunneling events on the contrary dissipate and thus heat up the Demon.}
    \label{fig1}
\end{figure*}

\section*{Results}
\paragraph*{\textbf{Model}}

Figure \ref{fig1}~(a) shows a schematic of the device. A metallic island is connected to two external leads via tunnel junctions, both with an equal tunneling resistance of $R_L = R_R = R$, where the indices refer to 'left' and the 'right' junctions. This forms the SET system that is measured. A detector - the actual Maxwell's demon is a single electron box, consisting of a metallic island connected to a grounded lead by a tunnel junction with tunneling resistance $R_D$. The system and the Demon islands are capacitively coupled to each other, and the whole setup is coupled to a phonon bath at inverse temperature $\beta=1/(k_B T)$. Finally, the system is biased by voltage $V$ so that the current runs from left to right, and the total Hamiltonian is given by
\begin{equation} \begin{split}
\label{eq:Hamiltonian}
H&= \frac{eV}{2} l + \frac{eV}{2}(-l-x) +  E_C^{sys}(x-\lambda_x)^2\\
&+ E_C^{dem}(y-\lambda_y)^2 +\kappa (x-\lambda_x)(y-\lambda_y),
\end{split}\end{equation}
where $E_C^{sys}$ and $E_C^{dem}$ denote the charging energies of the system and the Demon island, respectively, $\lambda_x$ and $\lambda_y$ are external electrostatic control parameters, $x$ and $y$ denote the number of excess electrons in the system and the Demon, respectively, $l$ is the number of electrons on the left lead, and $\kappa$ is the coupling energy. The dynamics are bipartite meaning that state $(l, x, y)$ may change by consecutive single electron tunneling events 
through the left junction $(l, x, y) \to (l \pm 1, x \mp 1, y)$, the right junction $(l, x, y) \to (l, x \pm 1, y)$, or the Demon junction $(l, x, y) \to (l, x, y \pm 1)$. Each tunneling event $\itof$, as a short notation of $(l_i, x_i, y_i) \to (l_f, x_f, y_f)$, has an energy cost directly given by Eq. \eqref{eq:Hamiltonian} as $\Eitof = H(l_f, x_f, y_f) - H(l_i, x_i, y_i)$, and the corresponding tunneling rate is given by
\begin{equation}
\label{eq:rates}
\Ritof =\frac{1}{e^2 R_{\upsilon}} \frac{ \Eitof}{e^{\beta \Eitof}-1},
\end{equation}
where $\upsilon = L, R, D$ refers to the junction associated with the transition $\itof$ (cf. Fig. 1). Higher order tunneling events are neglected, which is justified when tunneling resistances are much higher than the quantum resistance, i.e. $R, R_D \gg R_K = h / e^2$.

\paragraph*{\textbf{Energetics of electron tunneling in the setup}}

Next, we consider the operation of the setup at $\lambda_x=\lambda_y= {1}/{2}$, $eV < \kappa$, and $k_B T \ll \kappa, E_C^{sys}, E_C^{dem}$. It is then sufficient to consider only the lowest energy states $(x,y)\in \{(0,0),(0,1),(1,0),(1,1)\}$. The energy cost for a tunneling event in the system is
\begin{equation} 
\label{eq:xEnergyCost} 
\ExoiLR = \kappa(y - \frac{1}{2}) \varmp \frac{eV}{2} = -\ExioLR,
\end{equation}
where the $+$ and $-$ signs are used in case of tunnelling through the left (L) or right junction (R), respectively, as indicated in the superscript on the left of $E$. The energy cost for a Demon tunneling event is
\begin{equation} 
\label{eq:yEnergyCost} 
\EyoiD = \kappa (x - \frac{1}{2}) = - \EyioD, 
\end{equation}
where $D$ denotes for the Demon. Note that neither Eq. \eqref{eq:xEnergyCost} nor \eqref{eq:yEnergyCost} depend on $l$.
The energy is minimized when the islands have a single excess electron in total. Escaping the corresponding states $(0, 1)$ and $(1, 0)$ has 
an energy cost $ \kappa / 2$  for the Demon, 
and $(\kappa - eV) / 2$ for the system.   Relaxing back from $(1, 1)$ or $(0, 0)$ has an energy cost $- \kappa / 2$ for the Demon, and $ -(\kappa + eV)/ 2$ for the system. With an appropriate choice of $R_D \ll R$ and $V$, it is possible to realize a situation, where the energetically unfavored states $(1, 1)$ and $(0, 0)$ tend to relax through the Demon tunnel junction. As a result, when a tunneling event occurs in the system, cooling it by $(\kappa - eV) / 2$, the Demon rapidly reacts through another tunneling event, resuming the setup back to its ground state. This forms a cycle, illustrated in Fig. \ref{fig1} (b), where electric current flows through the SET while cooling it down by $\kappa - eV$ for each passing electron apparently violating Joule's law \cite{Koski2015}. However, Joule's law is retained by noting the heat $\kappa$ dissipated in the Demon.

\paragraph*{\textbf{Thermodynamics of the Demon}}

The probability distribution of the state $(l_i,x_i,y_i)$, $p_i \equiv p_{l_i x_i y_i}$, follows the master equation
$\dot p_i = -\sum_f \Jitof,$
where
\begin{equation}
\Jitof = \Ritof p_i - \Rftoi p_f,
\label{eq:Current}
\end{equation}
is the particle current from $(l_i,x_i,y_i)$ to $(l_f,x_f,y_f)$. We are interested in performance of the setup at steady state $\dot p_{l, x, y} = 0$. Such a state has no knowledge on the actual value of number of electrons on the left lead, $l$, i.e. $p_{l, x, y} = p_0 p_{x, y}$. 
The total entropy $S_{\rm tot}$ is a sum of the (dimensionless) Shannon entropy $S = -\sum_k p_k \ln p_k$ and the reservoir entropy $S_{\rm r} = \beta Q_T$ \cite{Schnakenberg1976}. The entropy production rate can be expressed as
\begin{equation} 
\dot S_{\rm tot} = \frac{1}{2} \sum_{i, f} \Jitof \ln\left(\frac{p_i \Ritof}{p_f \Rftoi}\right), 
\label{eq:EntropyProduction}
\end{equation}
which is always non-negative.
Further, proceeding as proposed in Ref. \cite{Horowitz2014}, Eq. \eqref{eq:EntropyProduction} splits in two non-negative contributions: One produced by tunneling events in the system, 
\begin{equation}
\begin{aligned}
\label{eq:SystemEntropy}
\dot{S}_{\rm tot}^X
= &\sum_{ x, y}  \JxpL \ln \left(\frac{\VxpL{\Rate} p_{ x, y}}{\VxmL{\Rate} p_{x+1, y}}\right)\\ 
+ &\sum_{ x, y} \JxpR \ln \left(\frac{\VxpR{\Rate} p_{x, y}}{\VxmR{\Rate} p_{ x+1, y}}\right) \\
= &\beta \dot Q_S - \dot I^X \geq 0,\\
\end{aligned}
\end{equation}
and another describing entropy produced by tunneling events in the Demon:
\begin{equation}
\begin{aligned}
\label{eq:DetectorEntropy}
\dot{S}_{\rm tot}^Y
= &\sum_{x, y}  \JypD \ln \left(\frac{\VypD{\Rate} p_{x, y}}{\VymD{\Rate} p_{x, y+1}}\right)\\ 
= &\beta \dot Q_D - \dot I^Y \geq 0,\\
\end{aligned}
\end{equation}
where $\dot I^{X}$ and $\dot I^{Y}$ are the changes in the mutual information $I=\ln [ p_{x,y} /(\sum_x p_{x,y} \sum_y p_{x,y}) ]$ due to the tunneling events in the Demon and the system, respectively, and $\dot Q_S=\dot Q_L+\dot Q_R$ and $\dot Q_D$ are the heat dissipation rates in the system and the Demon. The heat dissipation rate in each junction is
\begin{equation} \begin{split}
\label{eq:HeatDissipation}
 \dot Q_{L (R)} &= - \sum_{x, y} \VxpLR{\Current} \VxpLR{\Energy}; \\ 
\dot Q_D &= -\sum_{x, y} \VypD{\Current}\VypD{\Energy}.
\end{split} \end{equation}
The substitution with $\dot Q$ as in Eqs. \eqref{eq:SystemEntropy} and \eqref{eq:DetectorEntropy} results from local detailed balance, $Q_{i \to f}=-\Eitof=k_BT_\upsilon\ln (\Ritof / \Rftoi)$ \cite{Seifert2008}. The term
\begin{equation}
\dot I^Y=\sum_{x,y} W^{y \to y+1}_{x} \ln{\frac{p(x|y+1)}{p(x|y)}},
\end{equation}
where $W^y_{x_i \to x_f}={}^L \Gamma^y_{x_i \to x_f}+{}^R \Gamma^y_{x_i \to x_f}$. The term $\dot I^Y$ is the rate of mutual information produced by the Demon and quantifies how much transitions in $y$ increase correlation between $x$ and $y$ \cite{Parrondo2015}. In steady state the total time derivative of $I$ vanishes, but there is a flow of information $\dot I^Y=-\dot I^X$ between the Demon and the system. The terms $\dot I^{X}$  and $\dot I^{Y}$ also give the change in the Shannon entropy of the total system induced by a transition in the system and the Demon, respectively.

\paragraph*{\textbf{Demon as a refrigerator}}

\begin{figure*}{}
    \includegraphics[width=0.8\textwidth]{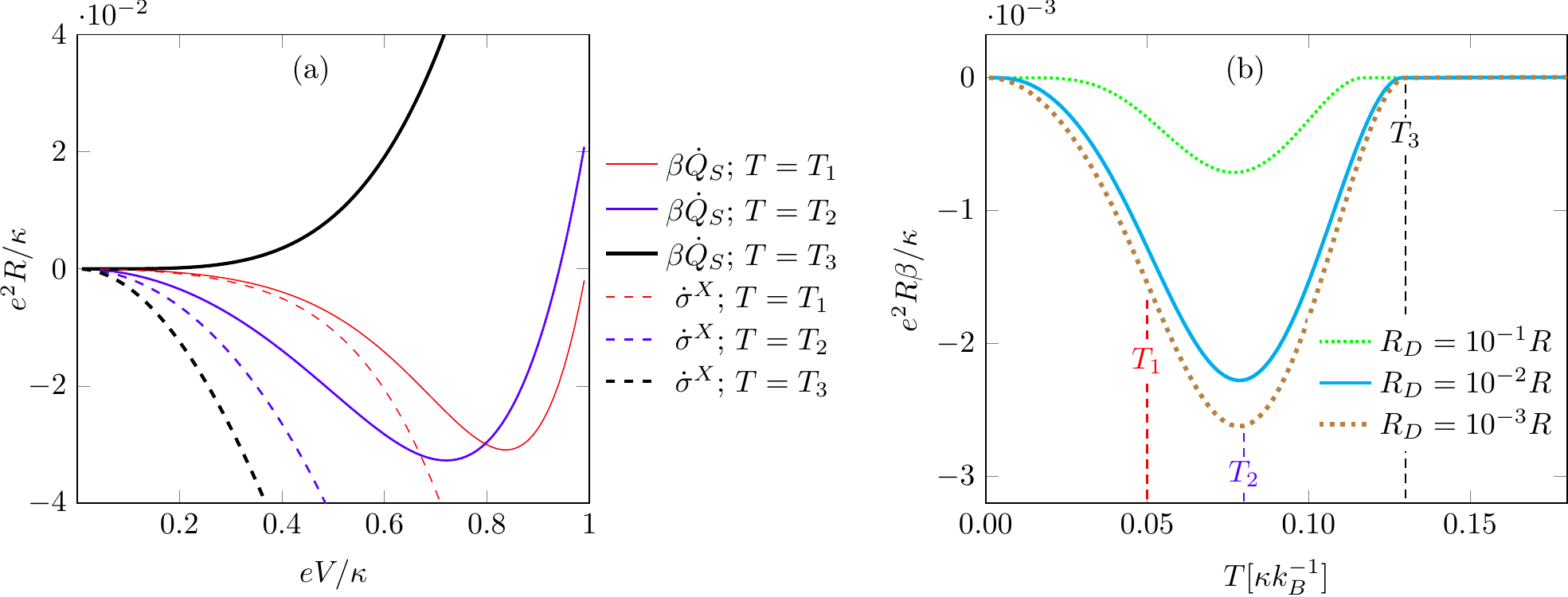}
    \caption{\textbf{Entropy production rate and cooling power dependence on temperature and bias voltage.} (a): Entropy production rate $\beta \dot{Q}_S$ and the coarse grained entropy production rate $\dot \sigma^X$ in the fast Demon limit ($R_D=10^{-3} R$) in different operating temperatures as a function of bias to coupling energy ratio. The coarse grained entropy is always negative and underestimates the entropy production. At low enough operating temperatures, there exists an optimal non zero bias voltage where the cooling is maximized. In higher temperatures no cooling is obtained. Temperatures used here are $T_1=0.05 \kappa k_B^{-1}$, $T_2=0.08 \kappa k_B^{-1}$ and $T_3=0.13 \kappa k_B^{-1}$. (b): Minimum system dissipation rate $\dot Q_S$ (with optimal bias voltage) as a function of operating temperature with three different Demon reaction rates ($R_D^{-1}$). Smaller resistance $R_D$ makes the Demon faster and more cooling is obtained. At temperatures higher than $T_3$ no cooling is obtained, while there exists an optimal operating temperature $T_2$ where the cooling power is maximized. Results are obtained by numerically solving the master equation with rates of Eq. \eqref{eq:rates}.}
\label{fig2}
\end{figure*}

In the low temperature regime, where both the system and the Demon have only two possible values of charge occupancy, the probability distribution is given by
\begin{equation} \begin{split} 
\label{eq:SteadyState}
p_{0, 1} &= p_{1, 0} = \frac{1}{2} \frac{\Rate_r}{\Rate_r + \Rate_e}; ~~ p_{0, 0} = p_{1, 1} = \frac{1}{2} \frac{\Rate_e}{\Rate_r + \Rate_e},
\end{split} \end{equation}
where $\Rate_r = \RxoiyoL + \RxoiyoR + \RxoyoiD$ is the relaxation rate and $\Rate_e = \RxoiyiL + \RxoiyiR + \RxoyioD$ is the excitation rate. 
For any $V \neq 0$, $\dot I^Y > 0$, implying that the tunneling events over the Demon junction on average increase the correlation between $x$ and $y$. Since $\dot I^X = -\dot I^Y$, the mutual information produced by the Demon is consumed in the system. To satisfy Eq. \eqref{eq:DetectorEntropy} the Demon must dissipate enough heat to its environment. The negative flow of information $\dot I^X$ allows for negative  $\beta \dot Q_S<0$ dissipation rate for the system without breaking the second law of Eq. \eqref{eq:SystemEntropy}, as shown in Fig. \ref{fig2} (a).

The heat dissipation rate in the system, Eq. \eqref{eq:HeatDissipation}, may be written as:
\begin{equation}
\begin{aligned}
&\dot Q_L = \dot Q_R = -\left(\frac{\kappa - eV}{2}\Gamma_{x: 1 \to 0}^{y = 0} + \frac{\kappa + eV}{2}\Gamma_{x: 0 \to 1}^{y=1}\right) p_{0, 1} \\
&+ \left(\frac{\kappa - eV}{2}\Gamma_{x: 0 \to 1}^{y = 0} + \frac{\kappa + eV}{2}\Gamma_{x: 1 \to 0}^{y=1}\right) p_{1, 1}, 
\end{aligned}
\end{equation}
where the first term is always negative, and the second term is always positive. Thus increasing the probability $p_{0, 1}$ increases the cooling power. Therefore, as can be seen from Eq. \eqref{eq:SteadyState}, the maximum cooling power is obtained when the tunneling rate over the Demon junction is maximized \cite{Koski2015}. This is in agreement with the numerical results which show that a faster Demon ($R_D<R$) gives rise to more cooling power as shown in Fig. \ref{fig2} (b). The operating temperature $T$ has to be sufficiently low, less than $0.13 k_B^{-1} \kappa$, in order to obtain cooling. In addition, if $R_D<R$, the optimal temperature, where the cooling power is maximized is roughly at $0.08 k_B^{-1} \kappa$.

\paragraph*{\textbf{Coarse grained entropy}}

We next examine entropy production in the setup, but now assuming that only the states of the system and the Demon, $x$ and $y$, are observed, and focus on the information exchange between the system and the Demon similar to Refs. \cite{Horowitz2014, Strasberg2013}. Therefore, we only consider the change $x_i \to x_f$ but do not distinguish whether the electron tunnels through the left or the right junction. With this approach the total entropy production rate is again given by Eq. \eqref{eq:EntropyProduction}, but the $x$ degree of freedom changes at the effective rate $W^y_{x_i \to x_f}={}^L \Gamma^y_{x_i \to x_f}+{}^R \Gamma^y_{x_i \to x_f}$. The total entropy production rate of the system is (cf. Eq. \eqref{eq:SystemEntropy})
\begin{equation}
{S}_{\rm cg}^X
=\dot \sigma^X+\dot{I}^Y \geq 0,
\label{eq:Scgsplit}
\end{equation}
where $\dot \sigma^X =1/2 \sum_{i,f}J_{i \to f}\sigma^X$ and  $\sigma^{X}=\ln (W^y_{x_i \to x_f} /W^y_{x_f \to x_i})$ defines the (coarse grained) entropy produced by the transition $x_i \to x_f$. In our setup, for non-zero bias, the entropy $\dot \sigma^X$ is always negative and thus the device works as a Maxwell's Demon, as shown in Fig. \ref{fig2} (a).

\paragraph*{\textbf{Efficiency of production and utilization of information}}

\begin{figure}{}
    \includegraphics[width=0.48\textwidth]{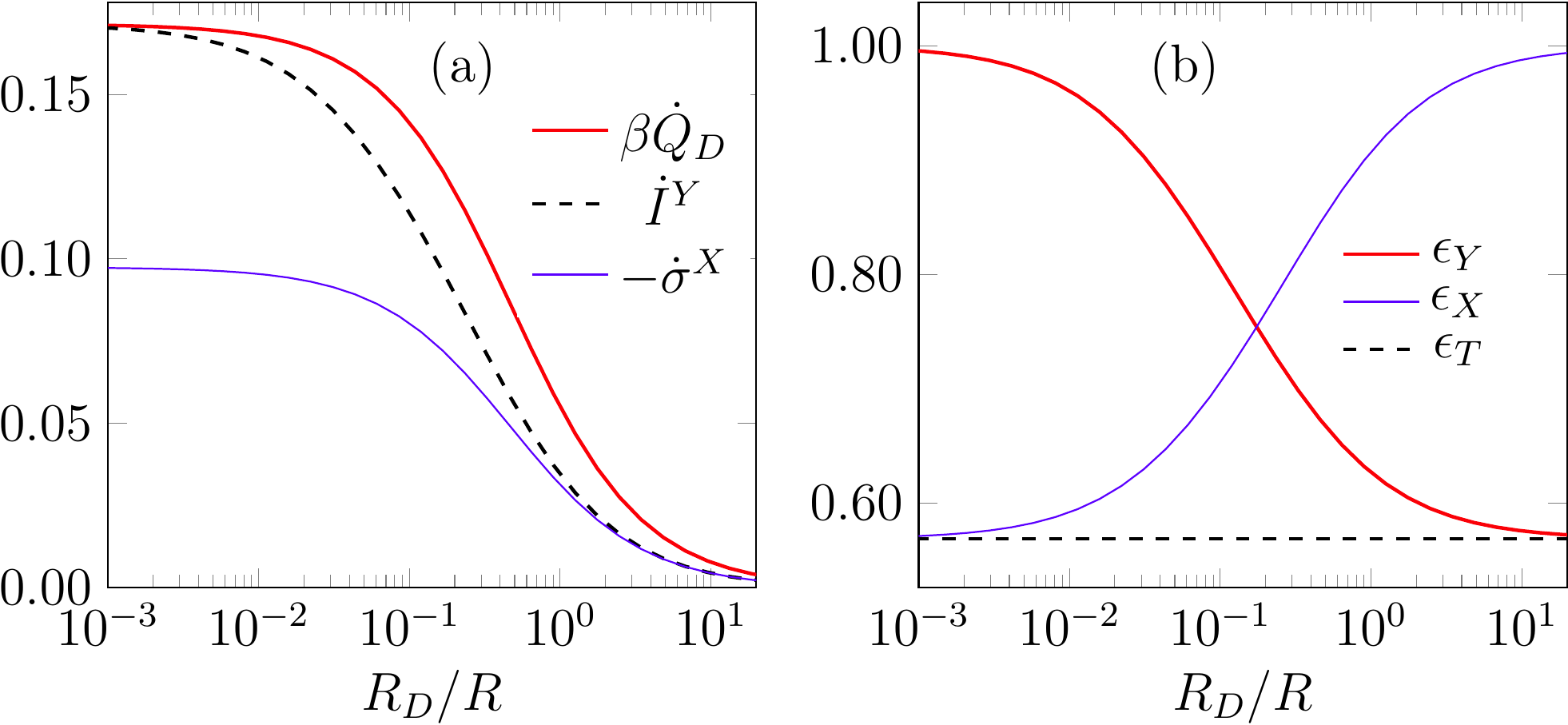}
    \caption{\textbf{Flow of information and the efficiency of its production and utilization.} (a): Entropy production rate in the Demon $\beta \dot Q_D$, flow of information $\dot I^Y$, and the coarse grained entropy production rate $\dot \sigma^X$ in the system as a function of Demon tunneling resistance ($R_D$). Smaller resistance makes the Demon faster. While the apparent entropy production rate in the system $\dot \sigma^X < 0$, the total entropy production rate $\dot {S}_{\rm cg}^X=\dot \sigma^X+\dot{I}^Y \geq 0$ (Eq. \eqref{eq:Scgsplit}). In addition, the Demon entropy production rate is always the largest of the three ensuring the inequality $\dot{S}_{tot}^Y= \beta \dot Q_D + \dot I^Y \geq 0$ (Eq. \eqref{eq:DetectorEntropy}). (b): The efficiency of information production, $\epsilon^Y$, its utilization, $\epsilon^X$, and that of the whole production-utilization, $\epsilon_T$. In the fast Demon limit ($R_D<<R$), the flow of information in the Demon equals the heat dissipation rate ($\epsilon^Y=1$), while in the slow limit the utilization of information flow becomes efficient ($\epsilon^X=1$). Parameters in both (a) and (b) are those optimal for maximum cooling power, $T=0.08\kappa k_B^{-1}$ and $eV/\kappa=0.72$, extracted from data shown in Fig. \ref{fig2} of the main text.}
    \label{fig3}
\end{figure}

 As shown in Fig. \ref{fig3} (a), a Demon with higher reaction rate ($R_D^{-1}$) is able to produce more information $\dot{I}^{Y}$. The entropic cost for sustaining the flow of information is the dissipation rate in the Demon $\beta \dot Q_D$ through heat \cite{Horowitz2014}. We define $\epsilon_Y=\dot{I}^Y / \beta \dot Q_D$ that characterizes the efficiency of the Demon information production. In Fig. \ref{fig3} (b) we show that a faster Demon is more efficient and in the limit of extremely fast reacting Demon, the flow of information $\dot I ^Y$ coincides with the heat dissipation rate, i.e. $\dot{I}^Y = \beta \dot Q_D$, corresponding the maximum efficiency of $\epsilon_Y=1$. The same result is obtained analytically by assuming the Demon is fast enough to thermalize on a time scale faster than the transitions occur in the system.

On the system side the apparent violation of the second law ($\dot \sigma^X < 0$) is provided by the flow of information $\dot{I}^{Y}$, which the system is able to utilize with efficiency $\epsilon_X=-\dot \sigma^X / \dot{I}^Y$. Contrary to $\epsilon_Y$, $\epsilon_X$ increases when the Demon is slower (large $R_D$) as shown in Figs. \ref{fig3} (a) and (b). We obtain, both analytically and numerically, that in the case of a very slow Demon, we have $\dot{I}^Y=-\dot \sigma^X$, which corresponds to the maximum efficiency of $\epsilon_X=1$.

Furthermore, a straightforward calculation shows that the efficiency of the whole measurement-feedback cycle, defined as $\epsilon_T=\epsilon_X \epsilon_Y=-\dot \sigma^X / \beta \dot Q_D $ is given by 
\begin{equation}
\epsilon_T=2/(\beta \kappa) \sigma^X_r,
\end{equation}
where $\sigma^X_r=\ln{[{W^{0}_{0 \to 1}}/{W^{0}_{1 \to 0}}]}$ is the coarse grained entropy production in the relaxation from ($0,0$) to ($1,0$) or equivalently from ($1,1$) to ($0,1$). Furthermore, this efficiency is independent of the Demon reaction rate $R_D^{-1}$, and thus a better Demon performance decreases the efficiency $\epsilon_X$ of the system as shown in Fig. \ref{fig3} (b).

\paragraph*{\textbf{Relation between coarse grained and bare entropies}}

We next study the relation between the entropy production rate $\beta \dot Q_S$ and $\dot \sigma^X$. Because the rates $W$ do not satisfy local detailed balance condition, $\sigma^X$ differs from the entropy $\beta Q_S$. However, the entropies are related as
\begin{equation}
\langle e^{-\beta Q_S} \rangle =e^{-\sigma^X},
\label{eq:relent}
\end{equation}
where $\langle \rangle$ denotes averaging over the conditional probabilities $P_{L}={}^{L} \Gamma^{xx'}_{y}/({}^{L} \Gamma^{xx'}_{y}+{}^{R} \Gamma^{xx'}_{y})$ and $P_{R}={}^{R} \Gamma^{xx'}_{y}/({}^{R} \Gamma^{xx'}_{y}+{}^{L} \Gamma^{xx'}_{y})$ to tunnel over the left and right junctions, respectively. Furthermore, Eq. \eqref{eq:relent} results in an integral fluctuation theorem for the coarse graining cost
$S_{cg}=\beta Q_S-\sigma^X$:
\begin{equation}
\langle e^{-S_{\rm cg}} \rangle = 1,
\label{eq:relent2}
\end{equation}
which by using Jensen's inequality gives
\begin{equation}
\langle S_{\rm cg} \rangle \ \geq 0,
\label{eq:Scg}
\end{equation}
implying that the coarse grained entropy underestimates the bare entropy production. This can also be seen in Fig. \ref{fig2} (a), 
while in the small bias $eV/\kappa \ll 1$ and at low temperature $T$ the entropy production rates $\beta \dot{Q}_S$ and $\dot{\sigma}^X$ coincide. By observing only the $x$ degree of freedom there can be an apparent violation of the second law, $\dot{\sigma}^X <0$, even in the regime where the bare entropy production rate $\beta \dot{Q}_S$ is positive. However, as can also be seen in Fig. \ref{fig3} (a), the coarse grained entropy production rate including the information, $\dot S^X_{\rm cg}=\dot \sigma^X+\dot{I}^Y $ is positive (Eq. \eqref{eq:Scgsplit}). The positivity of the coarse graining cost, Eq. \eqref{eq:Scg}, then also ensures positivity of the entropy production rate $\dot{S}_{\rm tot}^X= \beta \dot Q_S + \dot S^X \geq 0$ (Eq. \eqref{eq:SystemEntropy}).

\section*{Discussion}

To summarize, we have analyzed entropy production and flow of information in the experimentally feasible isothermal nanoscale device described in Fig. \ref{fig1} (a). The setup works as a Maxwell's demon device, where both the system and the Demon can be identified and where the measurement and the feedback are performed internally by the on-chip Demon. We have shown that depending on which variables are accessible for measurement, different apparent negative entropy productions result, however, the second law of thermodynamics always holds for the total combined system. Nevertheless, the performance and efficiency of the device to function as a cooler can be analyzed and adjusted by using thermodynamics of information. Thus, we conclude that information thermodynamics can be used to construct nanoscale devices with desired thermodynamic properties, e.g. to design dissipation in the device.

Acknowledgements: This research has been supported by the Academy of Finland through its Centres of Excellence Programs (project nos. 251748 and 250280), the European Union Seventh Framework Programme INFERNOS (FP7/2007-2013) under grant agreement no. 308850, and the V\"ais\"al\"a Foundation. We wish to thank Jukka Pekola, Samu Suomela, Ivan Khaymovich and Takahiro Sagawa for useful comments.

\begin{widetext}

\section*{Supplementary material}

\subsection*{Mutual information flow $\dot{I}^Y$ in the fast and slow Demon limits}

Mutual information is defined as $I  = \sum_{x,y} p_{x,y} \ln [p_{x,y}/(p^X_x p^Y_y)]$, where $p^X_x=\sum_y p_{x,y}$ and $p^Y_y=\sum_x p_{x,y}$ are the marginal distribution functions of the system and the Demon, respectively. In a transition from $x,y'$ to $x,y$ the change of mutual information is given by
\begin{equation}
\Delta I_{x}^{y' \to y} =\ln {\frac{p_{x,y}p^X_x p^Y_{y'}}{p^X_{x} p^Y_y p_{x,y'}}}=\ln {\frac{p_{x,y} } { p_{x,y'}}},
\label{mutinjump}
\end{equation}
where we used the symmetry property $p^Y_0=p^Y_1=p^X_0=p^X_1=1/2$, which results from the fact that the energetically favoured states $(0,1)$ and $(1,0)$ are equally probable to be occupied as are the energetically unfavoured states $(1,1)$ and $(0,0)$.

\subsubsection*{Fast Demon}

By assuming that the Demon is fast enough to thermalize between the transitions in system state $x$, the conditional probability for state $y$ given that the state of the system is $x$, $p(y|x)$, is given by
\begin{equation}
p(y|x)= \frac{{}^D\Gamma_{x}^{y' \to y}}{\gamma^D},
\label{eq:cond}
\end{equation}
where $\gamma^D={}^D\Gamma_{x}^{y \to y'}+{}^D\Gamma_{x}^{y' \to y}$. We note that $\gamma^D$ is independent of $x$, i.e. ${}^D\Gamma_{x=1}^{y \to y'}+{}^D\Gamma_{x=1}^{y' \to y}={}^D \Gamma_{x=0}^{y \to y'}+{}^D\Gamma_{x=0}^{y' \to y}$. The joint distribution function $p_{x,y}$ can be then written as 
\begin{equation}
p_{x,y}=p^X_x p(y|x)=\frac{1}{2} \times \frac{{}^D\Gamma_{x}^{y' \to y}}{\gamma^D}.
\label{pfast}
\end{equation}
By inserting the joint distribution function of Eq. \eqref{pfast} to Eq. \eqref{mutinjump} we obtain:
\begin{equation}
\Delta I_{x}^{y' \to y} =\ln {\frac{{}^D\Gamma_{x}^{y' \to y}}{{}^D\Gamma_{x}^{y \to y'}}}
\end{equation}

Thus the mutual information flow $\dot{I}^Y$ is given by
\begin{equation}
\dot{I}^Y=\sum_{y \geq y' ; x} J_{x}^{y' \to y} \Delta I_{x}^{y' \to y} =\sum_{y \geq y' ; x} J_{x}^{y' \to y} \ln {\frac{{}^D\Gamma_{x}^{y' \to y}}{{}^D\Gamma_{x}^{y \to y'}}}
=\beta \dot Q_D,
\end{equation}
where we used the local detailed balance condition for rates ${}^D\Gamma$.

\subsubsection*{Slow Demon}

By assuming that in the slow Demon limit the system thermalizes we obtain
\begin{equation}
p_{x,y}=p^Y_y p(x|y)=\frac{1}{2} \times \frac{W_{x' \to x}^{y}}{\gamma^X},
\label{pslow}
\end{equation}
where $\gamma^X=W_{x' \to x}^{y}+W_{x \to x'}^{y}$ and rate $W$ was defined in the main text as the sum rate $W={}^L \Gamma+{}^R \Gamma$ at which the $x$ degree of freedom changes. Similarly as in the fast Demon case, by inserting the joint distribution function of Eq. \eqref{pslow} to Eq. (\ref{mutinjump}) and using the symmetry relation $p^X_0=p^X_1$ we obtain:
\begin{equation}
\Delta I_{x}^{y' \to y} =\ln {\frac{W_{x' \to x}^{y}}{W_{x \to x'}^{y}}}.
\end{equation}
The mutual information flow is then given by
\begin{equation}
\dot{I}^Y=\sum_{y \geq y' ; x} J_{x}^{y' \to y} \Delta I_{x}^{y' \to y} =\sum_{y \geq y' ; x} J_{x}^{y' \to y} \ln {\frac{W_{x' \to x}^{y}}{W_{x \to x'}^{y}}}
\label{mutinfslow}
\end{equation}
Because we operate in the steady state, currents in and out from a state $(x,y)$ must be equal. Thus $J_{x=0}^{y:0 \to 1}=J_{x:0 \to 1}^{y=1}=J_{x=1}^{y:1 \to 0}=J_{x:1 \to 0}^{y=0}$. Furthermore, $J_{x}^{y' \to y}=-J_{x}^{y \to y'}$ and therefore $J_{x}^{y' \to y}=-J_{x'}^{y' \to y}$, if $x' \neq x$. By using these relations the mutual information flow of Eq. \eqref{mutinfslow} is given by
\begin{equation}
\dot{I}^Y=-\sum_{x \geq x' ; y} J_{x' \to x}^{y} \ln {\frac{W_{x' \to x}^{y}}{W_{x \to x'}^{y}}}=-\dot \sigma^X
\end{equation}

\subsection*{Derivation of the efficiency of the measurement-feedback cycle $\epsilon_T$}

Since the energy cost for tunnelings events in the system are related as ${}^{L(R)}E^{y=0}_{x:0 \to 1}={}^{R(L)}E^{y=1}_{x:1 \to 0}$, as can seen from Eq. (3) of the main text, the rates $W$ satisfy a relation $W_{x:0 \to 1}^{y=0}=W_{x:1 \to 0}^{y=1}$ and $W_{x:0 \to 1}^{y=1}=W_{x:1 \to 0}^{y=0}$. By using these relations and the fact that $J_{x:0 \to 1}^{y=0}=-J_{x:0 \to 1}^{y=1}$, we obtain
\begin{equation}
\dot\sigma^X=\sum_{x \geq x' ; y} J_{x' \to x}^{y} \ln {\frac{W_{x' \to x}^{y}}{W_{x \to x'}^{y}}}
=J_{x:0 \to 1}^{y=0}\ln {\frac{W_{0 \to 1}^{0}}{W_{1 \to 0}^{0}}}
+J_{x:0 \to 1}^{y=1}\ln {\frac{W_{0 \to 1}^{1}}{W_{1 \to 0}^{1}}}
=2 J_{x:0 \to 1}^{y=0}\ln {\frac{W_{0 \to 1}^{0}}{W_{1 \to 0}^{0}}}
\label{appsigmax}
\end{equation}
The entropy production rate in the Demon is given by 
\begin{equation}
\beta \dot Q_D =\sum_{y \geq y' ; x} J_{x}^{y' \to y} \ln {\frac{{}^D\Gamma_{x}^{y' \to y}}{{}^D\Gamma_{x}^{y \to y'}}}
=J_{x=0}^{y:0 \to 1} \ln {\frac{{}^D\Gamma_{x=0}^{y:0 \to 1}}{{}^D\Gamma_{x=0}^{y:1 \to 0}}}
+J_{x=1}^{y:0 \to 1} \ln {\frac{{}^D\Gamma_{x=1}^{y:0 \to 1}}{{}^D\Gamma_{x=1}^{y:1 \to 0}}}
=J_{x=0}^{y:0 \to 1} \beta \kappa,
\label{appbetaqd}
\end{equation}
where we used the fact that $J_{x=0}^{y:0 \to 1}=-J_{x=1}^{y:0 \to 1}$ and local detailed balance condition of rate ${}^D\Gamma$. Thus, combining Eqs. \eqref{appsigmax} and \eqref{appbetaqd} and using $J_{x=0}^{y:0 \to 1}=-J_{x:0 \to 1}^{y=0}$ we obtain
\begin{equation}
\frac{-\sigma^X}{\beta \dot Q_D}
=\frac{2}{ \beta \kappa}\sigma^X_r,
\end{equation}
where $\sigma^X_r=\ln{[{W^{0}_{0 \to 1}}/{W^{0}_{1 \to 0}}]}$.

\subsection*{Derivation of Eqs. (15) and (16) of the main text}

In a system tunneling event from $(x',y)$ to $(x,y)$ the coarse grained entropy production is given by
\begin{equation}
\sigma^X=\ln{\frac{{}^L \Gamma_{x' \to x}^{y}+{}^R \Gamma_{x' \to x }^{y}}{{}^L \Gamma_{x \to x' }^{y}+{}^R \Gamma_{x \to x'}^{y}}}
\end{equation}
The conditional probability to tunnel over the left/right junction is given by
\begin{equation}
P_{L/R}=\frac{{}^{L/R} \Gamma_{x' \to x}^{y}}{{}^{L/R} \Gamma_{x' \to x}^{y}+{}^{R/L} \Gamma_{x' \to x}^{y}}.
\end{equation}
Thus
\begin{equation}
P_Le^{-\beta {Q}_{L}}+P_Le^{-\beta {Q}_{R}}
=P_L \exp{[-\ln{\frac{{}^{L} \Gamma^{xx'}_{y}}{{}^{L} \Gamma^{x'x}_{y}}}]}
+P_R \exp{[-\ln{\frac{{}^{R} \Gamma^{xx'}_{y}}{{}^{R} \Gamma^{x'x}_{y}}}]}
=e^{-\sigma^X}.
\end{equation}

The average $\langle e^{-\beta Q_S} \rangle$ over a process of fixed amount time is given by
\begin{equation}
\langle e^{-\beta Q_S} \rangle = \langle e^{-\beta {Q}_{L}} \xi^L+e^{-\beta {Q}_{R}}\xi^R  \rangle
= \langle e^{-\sigma^X} \rangle,
\end{equation}
where $\xi^{L/R}$ is the indicator function giving $\xi^{L/R}=1$ if the tunneling is over the left/right junction and $0$ otherwise. The equation above can be written as
\begin{equation}
\langle e^{-(\beta Q_S-\sigma^X)} \rangle = \langle e^{-S_{cg}} \rangle= 1,
\end{equation}
where $S_{cg}$ is the coarse graining cost.

\end{widetext}

\end{document}